\documentclass[a4paper,10pt]{article}
\usepackage{amssymb}
\usepackage{amsmath}
\usepackage{latexsym}
\usepackage{array}
\usepackage{graphicx}

\usepackage{theorem}
\theoremstyle{break}

\def\qed{\hfill\hbox{\rule[-2pt]{3pt}{6pt}}}

\def\Z{{\mathbb Z}}

\newtheorem{defi}{Definition}[section]
\newtheorem{theo}{Theorem}[section]
\newtheorem{prof}{Proof}[section]

\begin{document}
\title{A property of the rule 150 elementary cellular automaton [2014/01/20]}
\author{Yuji Kaneko \\ \\
Graduate school of Mathematical Sciences, \\
the University of Tokyo, 3-8-1 Komaba, Tokyo 153-8914, Japan}

\date{}

\maketitle

\begin{abstract}
We studied the rule 150 elementary cellular automaton in terms of the distribution of the spacings of the singular values of 
the matieces obtained from proper time evolutions patterns.
The distribution has strong resembrance to that of the random matrices which is derived from Painlev\'e V equation.
Some analytic results for the relative period of the ECS are also presented.
\end{abstract}

\section{Introduction}
\label{sec1}
\subsection{Elementary cellular automaton}
The relations between random matrices and the Painlev\'e equations are well known. The distribution of largest eigenvalue of random matrices is derived 
from Painlev\'e II \cite{tr} equation and the distribution of nearest neighbor spacings for the eigenvalues is derived from Painlev\'e V \cite{jim}.
The relations between a random matrix and a cellular automaton (CA) have also been interested; the distribution of largest eigenvalues was investigated in the study of fluctuations of interfaces in which used a probabilistic CA \cite{tak}. A question is wheter we may have a similar relation in a deterministic CA. 

An elementary cellular automaton (ECA) is a one-dimensional, two-valued, 
three-neighborhood deterministic dynamical system \cite{wo}.
There are 256 kinds of ECAs and each ECA is numvered by a nonnegative integer from $0$ to $255$ according to its time evolution rule.
Let $u_n^t\in \{0,\, 1\}$ be the value of the $n$th cell at time step $t\in {\mathbb{Z}}_{\geq 0}$.
\begin{equation}
\label{e1}
u^{t+1}_{n}=u^{t}_{n-1}+u^{t}_{n}+u^{t}_{n+1} \pmod{2}
\end{equation}
The ECA determined by (\ref{e1}) is called the rule $150$ ECA.
We impose a periodic boundary condition: $u_n^{t}=u_{N+n}^{t}$, where $N\in \Z$ is the system size. 
We rewrite (\ref{e1}) as 
\begin{equation}
\label{e2}
{\bf{u}}^{t+1}= A_{150}{\bf{u}}^{t} \pmod{2}
\end{equation}
where ${\bf{u}}^{t}:={}^{t}(u_{1}^{t},\cdots,u_{N}^{t})$ is the state vector and $A_{150}$ is the $N \times N $ transition matrix.

The rule150 is time reversible under the condition of $N\equiv 1,2\pmod{3}$\cite{nobe}. 
We restrict ourselves to the system size $N$ to $1,2\pmod{3}$.
Then, there exists a positive integer $T({\bf{u}}^{0})\in\mathbb{N}$ such that
\begin{equation}
\label{e31}
{\bf{u}}^{0}=A_{150}^{T({\bf{u}}^{0})}{\bf{u}}^{0}
\end{equation}
for any initial state ${\bf{u}}^{0}$.
We call the minimum value of $T({\bf{u}}^{0})$ the fundamental period of ${\bf{u}}^{0}$.



\begin{figure}[htbp]
\centering
\includegraphics[width=120mm]{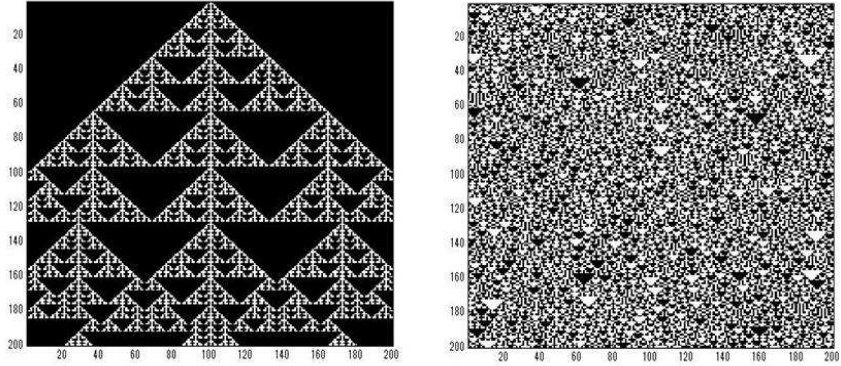}
 \label{fig:one}
\caption{Tow typical time evolution patterns of the rule150 ECA}
\end{figure}
The figures \ref{fig:one} show typical aspects(1=white, 0=black) of time evolution of the rule150.
The left figure corresponds to the initial state with single nonzero cell. 
The time evolution pattern exhibits a fractal-like pattern. 
On the other hand, the initial state of the right figure is a disordered state and its time evolution pattern and its time evolution pattern is somehow chaotic.

\begin{figure}[htbp]
\centering
\includegraphics[width=120mm]{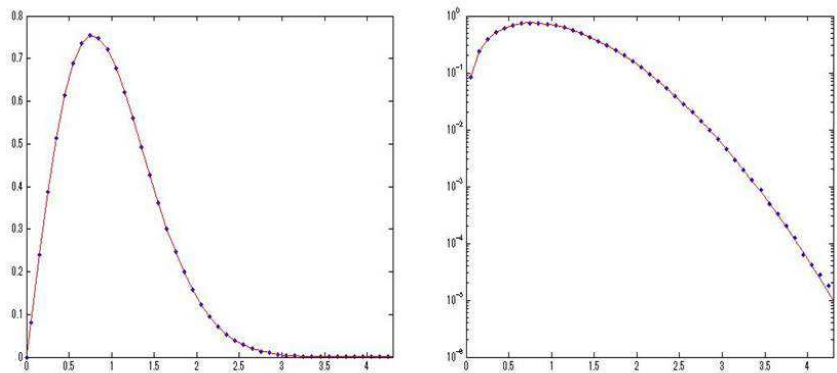}
 \label{fig:gobi}
 \caption{Distribution of spacings of singular values of the ECA rule150 time evolution pattern.} 
\end{figure}

Wolfram named Class 3 to the set of the rules which output patterns like the right figure of Fig.~\ref{fig:one}.
He mentioned that  \lq\lq Class 3 yields chaotic aperiodic patterns''.\cite{wo}
However, since the word \lq\lq chaotic'' is not well defined, 
the method to quantify the Wolframs' Class has been studied from the beginning of his proposal.
In our previous study \cite{kan}, we propose a method of the quantification in terms of distribution of eigenvalues of correlation matrices of time evolution patterns.
We characterize chaotic behavior of Class 3 by comparing output patterns with random patterns (random
matrix) by using the spacing distribution of singular values.
As a result, it became apparent that there is close agreement.
In particular, we found very good coincidence in the rule 150. 
We show the numerical results in Fig.~\ref{fig:gobi}.
In the left figure of Fig.~\ref{fig:gobi}, horizontal axis is spacing of eigenvalue and vertical axis is relative frequency of that spacing. The numerical result of the rule 150 with $N=1792$ is indicated by dots.
The rule150's singular values calculated from ${}^{t}U_cU_c$, here $U_c$ is the periodic time evolution pattern $U_c={}^{t}({\bf{u}}^{0},\cdots,{\bf{u}}^{T-1})$.
The solid line shows the theoretical results at $N\rightarrow\infty$ of random matrix (Gaussian Orthogonal Ensemble GOE) which is derived from the JMMS equation\cite{jim}. This equation is a kind of Painlev\'e V. The right figure is the semilog plot of the same figure.
%
%
%
%
The rule150's spacing distribution is closely coincident with the theoretical curve.
There is another example of  distribution which agrees with that of a random matrix, which is known as
the Montgomery-Odlyzko low. 
This low insists that the distribution of spacings between zeros of the Riemann zeta function closely agrees with that of eigenvalues of a random matrix \cite{me}.

The time evolution of ECA is deterministic, so it is non-trivial that the property of the matrix which comes from the rule150 is closely related to a  random matrix.
Thus theoretical investigation of the distribution of spacings of the rule150 is interesting.
If we do not require high accuracy, we can find similarity between a random matrix and the other ECAs of Class 3 than the rule150 ECA\cite{kan}\cite{kan2}.
However, other rules do not exhibit such accuracy of coincidence and it is also important to find the reason why the rule 150 is special. 

The main purpose of this paper is to show that the rule150 has special property in calculating the spacing distribution of a \lq\lq chaotic'' ECA.
We will also elucidate that the periodic time evolution matrix $U_c$ represents the behavior of CA in the long-time limit.

\section{Methods of simulation}\label{numerical_simulation}
%
%
The procedure to obtain the spacing distribution of singular values numerically is as follows.
Here ${\bf{3}}\sim{\bf{6}}$ are the processes of unfolding \cite{boh}.

\begin{description}
\item[{\bf 1}] 
Obtain periodic pattern $U_c$ which corresponds to random initial state, and calculate eigenvalues of ${}^{t}U_cU_c$ numerically.
\item[{\bf 2}] Sort the eigenvalues $\{ x_1,\dots,x_N \}(x_{i-1}\leq x_{i},2\leq i \leq N)$.
Keep the eigenvalues from $x_{N/4+1}$ to $x_{3N/4}$ and discards the other eigenvalues, because it is known that the property of eigenvalues which locate in the vicinity of the edges of eigenvalues are different from the others.
\item[{\bf 3}] Construct the cumulative distribution function for {\bf 2} and multiply $N/2$.
\item[{\bf 4}] Approximate {\bf 3} by a polynomial function $f(x)$ (smoothing).
\item[{\bf 5}] Obtain spacings $s_i=y_i-y_{i-1}$, here $y_i=f(x_i)$.
\item[{\bf 6}] Repeat {\bf 1}--{\bf 5} for randomly selected initial states.
Let $L$ be the number of spacings of eigenvalues of one matrix $U_c$, 
$s_i^{(j)}$ be the $i$th spacing of $j$th $U_c$, and $M$ be the number of random samplings. 
We construct the set $S:=\{s_i^{(j)}|i=1,2,\dots,L;j=1,2,\dots,M\}$ and finally obtain relative frequency 
$P(s)$.
\end{description}
Note that we decrease the number of usable spacings in some cases, that is, $L\leq [N/2]-1$, where $[\cdots]$ denotes the Gauss symbol. 
One of such cases is the case $T<N$. 
In this case we cannot use trivial zero eigenvalues and theirs spacings. 
There was no sample which shows $T<N$ in the numerical calculation of the rule150. Figure \ref{fig:gobi}  was obtained with $N=1792, M=10000$.


Now we consider how to estimate the $P(s)$ in the limit $N \rightarrow\infty$ by the methods given above.
We have to evaluate the average of (\ref{eq4}). 
We denote by $Q(s|U_c)$ the conditional distribution of spacing $s$ for a given $U_c$, and by $R(U_c)$ the occurrence probability of $U_c$.
\begin{equation}\label{eq4}
P(s)=\int Q(s|U_c)R(U_c)dU_c
\end{equation}
The random matrix which corresponds to ${}^{t}U_cU_c$ is the Wishart matrix ${}^{t}XX$.
Here $X$ is a $p\times q$ rectangular random matrix $(p\geq q)$. 
If we fix the ratio $Q=p/q$ and $p\rightarrow\infty, q\rightarrow\infty$, 
then the spacing distribution of eigenvalues of the Wishart matrix is the same as that of the GOE.
When we compare ${}^{t}U_cU_c$ with ${}^{t}XX$, the rule150 has good properties (I)$\sim$(III)
at $N=7\cdot2^m$. We give the proofs of these statements in the next section.
\begin{itemize}
\item[(I)] €The maximum fundamental period is $N$.

\vspace{3mm}
In the books by Wolfram \cite{wo} \cite{wo2}, there are results of numerical simulation to investigate behavior of $T$ associated with increase of $N$. The behavior of the rule 150 is fairly complicated as the other rules contained Class 3.
However we can prove that the maximum fundamental period of the rule 150 is exactly $N$ at $N=7\cdot 2^{m}$ ($m \in \Z_+$).
This is a similar property of the Wishart matrix and we should focus on the rules which have similar property with it.
Note that $T$ is not proportional to $N$ in general. For example, the rule 45 has a very long periods which is close to $2^N$ \cite{wo2}.
In this case, it is predicted that $Q':=T/N$ will diverge when $N\rightarrow\infty$.
\item[(II)] Almost all fundamental periods are equal to $N$.

\vspace{3mm}
We cannot know the number of $U_c$ that has maximum fundamental period from (I).
If fundamental periods of many periodic patterns are less than $N$, those sample produce trivial zero eigenvalues. To prevent the occurrence of this situation, the Wishart matrix has condition $p\geq q$. But we can't choose $T$ in the ECA. It is essential difference between the random matrix and the ECA.

For this point, (II) insists almost every periodic patterns have maximum fundamental period.
This statement drive $R(U_c)$ to constant in the (\ref{eq4}) when $N\rightarrow\infty$.
Thus we expect that $N=7\cdot2^{m}$ is very useful for calculation of (\ref{eq4}).
\item[(III)] The rule150 is reversible.

\vspace{3mm}
This is obvious from \cite{nobe}. But this property is very important.
From this property we can conclude that random sampling for initial states is equivalent to 
random sampling for periodic patterns with maximum fundamental period.

The number of independent rules of ECA is 88 \cite{wo}. The rules belong to Class 3 are the rules 18,22,30,45,60,90,105,106,122,129,146,150. 
The reversible rules in Class 3 are
 rule 45,105,150 \cite{nobe}.
The rule 45 is reversible at $N=1\pmod{2}$ and the rule 105, 150 is reversible at $N=1,2\pmod{3}$.
The rule 105's time evolution is written as follows.
It is similar to the rule 150, but is not linear.
\begin{equation}
u^{t+1}_{n}=u^{t}_{n-1}+u^{t}_{n}+u^{t}_{n+1}+1 \pmod{2}
\end{equation}
\end{itemize}

An important point to compare the distribution function for an ECA to that of a random matrix is 
that the time evolution pattern of the ECA does not have any symmetry such as translational symmetry.  
It is also important that the ECA has reversiblity because any state is not a decaying state and we can use all the states in an orbit.
Hence these features (I) $\sim$ (III) are of particular interest. 

%
\section{A theorem for periodicity of the rule 150}
In periodic boundary conditions, the transient matrix $A_{150}$ of the rule 150
is wrettn by the shift matrix $\Lambda$ and identity matrix $E_{l}$. 
Here $E_{l}$ is $l\times l$ identity matrix. In case of $l=N$, we omit index $l$.
\begin{equation}
\label{eq:eq23}
\Lambda
= 
\left(
  \begin{array}{c|c}
      0        &  1    \\ \hline
      E_{N-1}         & 0
  \end{array}
  \right)
\end{equation}
\begin{equation}
\label{eq:eq24}
A_{150}=E+\Lambda+{\Lambda}^{-1}
\end{equation}

\begin{theo}\label{th4.1}
Assume 
$N=T=2^m(2^k-1)$ and $N=1,2\pmod{3}$.
Then $(E+\Lambda+\Lambda^{-1})^T=E$.
\end{theo}

\begin{prof}
In general, it holds that
\begin{equation}
\label{eq:eq39}
(E+\Lambda+\Lambda^{-1})^{2^m}=E+\Lambda^{2^m}+\Lambda^{-2^m}
\end{equation}
Hence, $T+2^m=2^{k+m}$ and thus 
\begin{align*}
(E+\Lambda+\Lambda^{-1})^{T+2^m}\\
&=(E+\Lambda^{2^m}+\Lambda^{-2^m})^{T}\\
&=E+\Lambda^{2^{m+k}}+\Lambda^{-2^{m+k}}\\
&=E+\Lambda^{2^{m}}+\Lambda^{-2^{m}}=(E+\Lambda+\Lambda^{-1})^{2^m}
\end{align*}
When $N=1,2\pmod{3}$, $\, E+\Lambda+\Lambda^{-1}\,$ is an inverse matrix. 
So we have the theorem \ref{th4.1}.
\qed

\end{prof}

Hereafter we restrict the system size to $N=7\cdot 2^m$.

\begin{theo}
The total number of initial states which do not have the maximum fundamental period is $2^{4\cdot 2^n}$.
\end{theo}

\begin{prof}
In this proof, \lq\lq period'' dose not mean the fundamental period, and it involves multiples of the fundamental period. 
The patterns that have periods shorter than $T$ must have period $2^m$ and/or $7\cdot2^{m-1}$.
We consider three cases.

\begin{itemize}
\item[1.] The total number of the states which have period $2^m$.

We count the number of solutions to the following equation for ${\bf{u}}\in({\mathbb{Z}}_2)^T$.
\begin{equation}
\label{eq:eq40}
(E+\Lambda+\Lambda^{-1})^{2^m}{\bf{u}}={\bf{u}}
\end{equation}
From (\ref{eq:eq39}), $(E+\Lambda^{2^m}+\Lambda^{-2^m}){\bf{u}}={\bf{u}}$.
Let $r:=2^m$. In case the solutions have period $2^{m}$, $u_{n-r}+u_{n+r}\equiv 0\pmod{2}$ holds for all $n$
(We denote the index of $u$ with $\mod{T}$).
Thus, the solutions satisfy
\begin{equation}
{}^{\forall}n,\,u_{n}\equiv u_{n+2r}\nonumber
\end{equation}
On the other hand, from the relation ${\rm{GCM}}(2r,T)=r$ and the periodic boundary condition, the solutions satisfy the following condition.
\begin{equation}
{}^{\forall}n,\, u_{n}=u_{n+r}=u_{n+2r}=\cdots=u_{n+6r}\nonumber
\end{equation}
The number of those state is $2^{2^m}$.

\item[2.] The total number of the states which have period $7\cdot2^{m-1}$.\\
\begin{equation}
\label{eq:eq41}
(E+\Lambda+\Lambda^{-1})^{7\cdot2^{m-1}}{\bf{u}}={\bf{u}}
\end{equation}
From (\ref{eq:eq39}), $(E+\Lambda^{2^{m-1}}+\Lambda^{-2^{m-1}})^{7}{\bf{u}}={\bf{u}}$.
Let $z:=2^{m-1}$ and $\Gamma:=\Lambda^{z}$, we obtain 
\begin{equation}
\label{eq:eq42}
(\Gamma^{-6}+\Gamma^{-4}+\Gamma^{-3}+\Gamma^{-1}+E+\Gamma^{1}+\Gamma^{3}+
\Gamma^{4}+\Gamma^{6}){\bf{u}}={\bf{u}}
\end{equation}
Hence, we have only to count the number of the solutions to the following equation.
\begin{equation}
(\Gamma^{-6}+\Gamma^{-4}+\Gamma^{-3}+\Gamma^{-1}+\Gamma^{1}+\Gamma^{3}+
\Gamma^{4}+\Gamma^{6}){\bf{u}}={\bf{0}}\nonumber
\end{equation}
Specifically, 
\begin{equation}
{}^{\forall}n,\,u_{n-6z}+u_{n-4z}+u_{n-3z}+u_{n-z}+u_{n+z}+u_{n+3z}+u_{n+4z}+u_{n+6z}
\equiv0\nonumber
\end{equation}
This equation is closed under 14 variables $\{u_{n+kz}\}_{k=-6}^{7}$.
\begin{equation}
{\bf{u}}':={}^{t}(u_{n-6z},u_{n-5z},u_{n-4z},\cdots,u_{n+5z},u_{n+6z},u_{n+7z})\nonumber
\end{equation}
So we use new notation ${\bf{u}}'$ and consider the following simultaneous equations.
\begin{equation}
\label{eq:eq43}
(\Lambda^{-6}+\Lambda^{-4}+\Lambda^{-3}+\Lambda^{-1}+\Lambda^{1}+\Lambda^{3}+
\Lambda^{4}+\Lambda^{6}){\bf{u}}'={\bf{0}}
\end{equation}
Let $K$ be the number of solutions to \eqref{eq:eq43}.
Then the number of solutions of the original equation \eqref{eq:eq41} is $K^z$.
The rank of the matrix corresponding to the left hand side of \eqref{eq:eq43} is 6. 
Therefore $K=2^{14-6}=2^8$.
Consequently, the number of state is $(2^8)^s=2^{4\cdot2^m}$.

\item[3.] The total number of the states which have period $2^{m-1}$.

The G.C.D of $2^m$ and $7\cdot2^{m-1}$ is $z(=2^{m-1})$.
The states which have period $z$ satisfy
\begin{equation}
(E+\Lambda+\Lambda^{-1})^z{\bf{u}}=(E+\Lambda^{z}+\Lambda^{-z}){\bf{u}}={\bf{u}}\nonumber
\end{equation}
$\forall n,\,u_{n-z}+u_{n+z}\equiv0\pmod{2}$. Let $r=2^n$ and the following equation holds.
\begin{equation}
{}^{\forall}n,\, u_{n}=u_{n+r}=u_{n+2r}=\cdots=u_{n+6r}\nonumber
\end{equation}
The number of those states is $2^{2^m}$.
\end{itemize}

Therefore the total number of states which do not have the maximum fundamental periodic patterns is 
\[
2^{4\cdot2^m}+2^{2^m}-2^{2^m}=2^{4\cdot2^m}
\] 
\qed
\end{prof}

We can conclude that when $N=7\cdot2^m,\, m\rightarrow\infty$, almost all states of the rule 150 have the maximum fundamental period patterns from the above theorem.
But, we have to pay attention to symmetry of the $U_c$ when we calculate the distribution of spacings for eigenvalues.
In this paper we use periodic boundary conditions and our systems has translational symmetry in space.
According to our numerical methods, those symmetry changes the spacing distribution qualitatively \cite{kan}\cite{kan2}.
For example in case of the rule 90, we observe the states which satisfy  ${\bf{u}}^{T/2}=\Lambda^{N/2}{\bf{u}}^{0}$ at  $N=7\cdot2^m$ and
the spacing distribution of eigenvalues of ${}^{t}U_cU_c$ disagrees with that of GOE.
In the next section, we prove that the rule 150 have few states like this when $N=7\cdot2^m$.


\section{A theorem for relative period of the rule 150}

\begin{defi}
Let $T'({\bf{u}}^{0})\,$ be the minimum positive integer which satisfies
\begin{equation}
\label{e3}
\Lambda^{S({\bf{u}}^{0})}{\bf{u}}^{0}=A_{150}^{T'({\bf{u}}^{0})}{\bf{u}}^{0}
\end{equation}
for a certain positive integer $\, S({\bf{u}}^{0})\in\mathbb{N}$ and
for any initial state ${\bf{u}}^{0}$.
We call $T'=T'({\bf{u}}^{0})$ the fundamental relative period of ${\bf{u}}^{0}$, and name $S=S({\bf{u}}^{0})$ the shift number for this $T'$ $(1 \le S \le N)$. 
If $T=T'$, the shift number $S=N$.
\end{defi}
We consier only the case $N=7\cdot2^m$.
\begin{theo}
When $m\rightarrow\infty$, almost all states of rule150 have periodic patterns which have maximum fundamental relative periods $T'=7*2^m=T=N$.
\end{theo}

\begin{prof}
In this proof  \lq\lq relative period'' dose not mean the fundamental relative period, and it allows multiples of the fundamental relative period. 
The patterns that have relative periods smaller than $T$ must have relative periods $T'=2^m$ with $S=S_1:=2^m, S_2:=2\cdot2^m, S_3:=3\cdot2^m$ and/or $T'=7\cdot2^{m-1}$ with $S=S_4:=N/2$.
\begin{itemize}
\item[1.] The total number of states belongs to $T'=2^m, S=S_1$. \\
\begin{equation}
(E+\Lambda+\Lambda^{-1})^{2^m}{\bf{u}}=\Lambda^{2^m}{\bf{u}}\nonumber
\end{equation}
Thus we should solve $\Lambda^{2^m}{\bf{u}}={\bf{u}}$. 
Let $r:=2^n$ and the following equation holds.
\begin{equation}
\label{eq:eq44}
{}^{\forall}n,\,u_{n}=u_{n+r}=u_{n+2r}=\cdots=u_{n+6r}
\end{equation}
The number of those states is calculated as $2^{2^m}$.
Here, the states represented by (\ref{eq:eq44}), which have period $2^m$ in space.
We can regard those states as the systems with smaller system size $N=2^m$.
In this case, since $\Lambda^{N/2}=\Lambda^{-N/2}$, $A_{150}^{N/2}=E$ and $T=2^{m-1}$. 
The number of these states is very few as is proven in the previous section.  
Hence we need not consider this case.
\item[2.] The total number of states belongs to $T'=2^m, S=S_2$.
\begin{equation}
\label{eq:eq45}
(E+\Lambda+\Lambda^{-1})^{2^m}{\bf{u}}=\Lambda^{2\cdot2^m}{\bf{u}}
\end{equation}
The following equation holds for $t=2\cdot2^m$.
\begin{equation}
\label{eq:eq46}
(E+\Lambda+\Lambda^{-1})^{2\cdot2^m}{\bf{u}}=\Lambda^{4\cdot2^m}{\bf{u}}
\end{equation}
From 
(\ref{eq:eq45}) and (\ref{eq:eq46}), we obtain

\begin{equation}
\label{eq:eq47}
(\Lambda^{4\cdot2^m}+\Lambda^{-4\cdot2^m}){\bf{u}}=
(\Lambda^{2\cdot2^m}+\Lambda^{-2\cdot2^m}){\bf{u}}
\end{equation}
Adding {\bf{u}} to both sides of (\ref{eq:eq47}), we find
\begin{equation}
\label{eq:eq48}
A_{150}^{4\cdot2^m}{\bf{u}}=A_{150}^{2\cdot2^m}{\bf{u}}
\end{equation}
Equation (\ref{eq:eq48}) means that $T=2^m$. 
Hence we need not consider this case.
%
\item[3.] The total number of states belongs to $T'=2^m, S_3=3\cdot2^m$.
\\
For $t=2^m,2\cdot2^m,3\cdot2^m$, the following three equations hold.
\begin{gather}
(E+\Lambda+\Lambda^{-1})^{2\cdot2^m}{\bf{u}}=\Lambda^{6\cdot2^m}{\bf{u}}\label{eq:eq49}\\
(E+\Lambda+\Lambda^{-1})^{5\cdot2^m}{\bf{u}}=\Lambda^{2^m}{\bf{u}}\label{eq:eq50}\\
(E+\Lambda+\Lambda^{-1})^{6\cdot2^m}{\bf{u}}=\Lambda^{4\cdot2^m}{\bf{u}}\label{eq:eq51}
\end{gather}
From (\ref{eq:eq49}), (\ref{eq:eq50}) and (\ref{eq:eq51}), we obtain
\begin{equation}
\label{eq:eq52}
A_{150}^{2^m}{\bf{u}}=\Lambda^{4\cdot2^m}{\bf{u}}
\end{equation}
From (\ref{eq:eq52}) and $A_{150}^{2^m}{\bf{u}}=\Lambda^{3\cdot2^m}{\bf{u}}$, we can obtain the solutions which have period $2^{m}$ in space.
As the case of the $T'=2^m,\, S=2^m$, these states are very few and we need not consider in the limit $m \rightarrow \infty$.
\item[4.] The total number of states belongs to $T'=7\cdot2^{m-1}, S_4=N/2$.
\begin{equation}
\label{eq:eq53}
(E+\Lambda+\Lambda^{-1})^{7\cdot2^{m-1}}{\bf{u}}=\Lambda^{7\cdot2^{m-1}}{\bf{u}}
\end{equation}
Using the same arguments as in (\ref{eq:eq42}), we obtain
\begin{equation}
\label{eq:eq54}
(\Lambda^{-6}+\Lambda^{-4}+\Lambda^{-3}+\Lambda^{-1}+E+\Lambda^{1}+\Lambda^{3}+
\Lambda^{4}+\Lambda^{6}+\Lambda^{7}){\bf{u}}'={\bf{0}}
\end{equation}
The rank of matrix corresponding to the left hand side of \eqref{eq:eq54} is 7. 
The number of states is $2^{7\cdot2^{m-1}}$. 
\end{itemize}
Thus we can evaluate the upper bound of the number of the states which do not have the periodic patterns with maximum fundamental relative period
as $2^{4\cdot2^m}+2^{7\cdot2^{m-1}}$. Thus the ratio of the number of the states calculated above to that of all the states is
less than
\begin{equation}
\label{eq:eq55}
\frac{2^{4\cdot2^m}+2^{7\cdot2^{m-1}}}{2^{7\cdot2^m}}
\end{equation}
The above ratio becomes to zero for $m\rightarrow\infty$. 
Therefore almost all states of the rule 150 have the periodic patterns with maximum fundamental relative period $T'=N=7\cdot2^m$.
\qed
\end{prof}
\section{Discussion and concluding remarks}
The distribution of the spacings of the singular values derived from the periodic patterns of
a chaotic cellular automaton was shown numerically to closely coincide with that of eigenvalues of  random matrices.
In this paper,  we showed that the rule 150 ECA with a particular system sizes $N=7\cdot2^m$ has special features 
which are suitable for compariosn to random matrices. 
In this case, almost all periodic patterns are regarded as square matrices and  we can use eigenvalues instead of singular values for rectangular matrices.
This property satisfies the Wishart matrix's condition for $Q\geq1$.
The rigorous derivation of the rule150's spacing distribution is a problem we would like to address in the future.
 

\section*{Acknowledgements}
The author wishes to thank Professors T.Tokihiro, J.Mada and S.M.Nishigaki for useful comments.


\end{document}